\documentclass{article}

\usepackage{graphicx}
\usepackage{subfigure}
\usepackage{xcolor}
\usepackage{geometry}
\usepackage{caption}
\usepackage{amsmath}
\usepackage{comment}
\usepackage{hyperref}
\usepackage{algorithm, algpseudocode, algorithmicx}

\newtheorem{thm}{Theorem}[section]

\title{Generating Weakly Chordal Graphs from Arbitrary Graphs}

\author{Sudiksha Khanduja \\
	School of Computer Science \\
	University of Windsor \\
	Windsor, Canada \\
	\and
	Aayushi Srivastava \\
	School of Computer Science \\
	University of Windsor \\
	Windsor, Canada \\
	\and
	Md. Zamilur Rahman \\
	School of Computer Science \\
	University of Windsor \\
	Windsor, Canada \\
	\and Asish Mukhopadhyay \\
	School of Computer Science \\
	University of Windsor \\
	Windsor, Canada \\
	}
\date{}

\begin{document}
\maketitle{}

\begin{abstract}
We propose a scheme for generating a weakly chordal graph from a randomly generated input graph, $G = (V, E)$. We reduce $G$ to a chordal graph $H$ by adding fill-edges, using the minimum vertex degree heuristic. Since $H$ is necessarily a weakly chordal graph, we use an algorithm for deleting edges from a weakly chordal graph that preserves the weak chordality property of $H$. The edges that are candidates for deletion are the fill-edges that were inserted into $G$. In order to delete a maximal number of fill-edges, we maintain these in a queue. A fill-edge is removed from the front of the queue, which we then try to delete from $H$. 
If this violates the weak chordality property of $H$, we reinsert this edge at the back of the queue. This loop continues till no more fill-edges can be removed from $H$. Operationally, we implement this by defining a deletion round as one in which the edge at the back of the queue is at the front.
We stop when the size of the queue does not change over two successive deletion rounds and output $H$. 
\end{abstract}

\section{Introduction}\label{introduction}
\begin{figure}[htb]
	\centering
	\subfigure[{\em $G$} \label{Fig-WCGExample}]{\includegraphics[scale=.65]{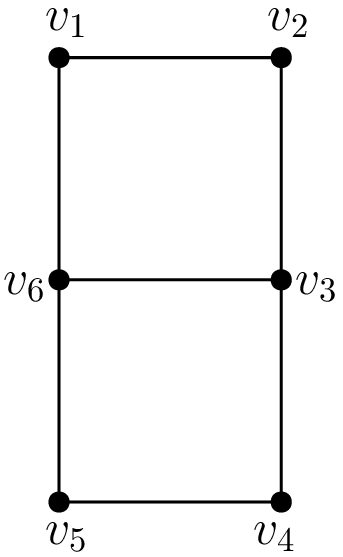}}\hspace{50pt}
	\subfigure[{\em $\overline{G}$} \label{Fig-WCGCompExample}]{\includegraphics[scale=.65]{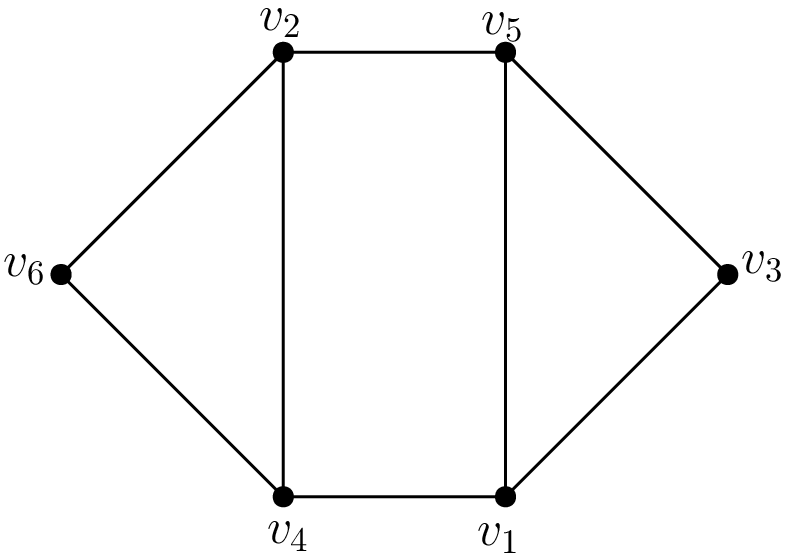}}
	\caption{{\em Weakly chordal graph~\cite{DBLP:journals/corr/abs-1906-01056}}}
	\label{Fig-WCGExamples}
\end{figure}

A graph $G = (V, E)$ is said to be weakly chordal if neither $G$ nor its complement, $\overline{G}$, has an induced chordless
cycle on five or more vertices (a hole). Figure~\ref{Fig-WCGExamples} shows an example of a weakly chordal graph, $G$, and its complement, $\overline{G}$. \\

 Weakly chordal graphs were introduced by Hayward in~\cite{DBLP:journals/jct/Hayward85} as a generalization of chordal 
graphs, who showed that these graphs form a subclass of the perfect graphs. An alternate definition that does not refer to the complement graph is that $G$ does not contain 
a hole or an anti-hole, which is the complement of a hole. Berry et al. \cite{DBLP:journals/njc/BerryBH00} gave a very different and interesting definition of a 
weakly chordal graph as one in wich every edge is $LB$-simplicial. They also proposed the open problem of generating a weakly chordal
graph from an arbitrary graph. A solution to this problem is the subject of this paper. \\

Early work on graph generation foucussed on creating catalogues of graphs of small sizes. Cameron et al. ~\cite{DBLP:journals/jgt/CameronCRW85}, for instancce, 
published a catalogue of all graphs on 10 vertices. The underlying motive was that such repositories were useful for providing
counterexamples to old conjectures and coming up with new ones. Subsequent focus shifted to generating graphs of arbitrary size, labeled and unlabeled, uniformly at random. As such a generation method, involved solving a counting 
problem, research was focused to classes of graphs for which the counting problem could be solved and yielded polynomial time generation algorithms. Among these were 
 graphs with prescribed degree sequence, regular graphs, special classes of graphs such as outerplanar graphs, maximal planar graphs. See~\cite{Tinhofer1990} for a survey work prior to 1990. \\
 
 As stated in~\cite{DBLP:journals/corr/abs-1906-01056}, there are many situations where we would like to generate instances of these
 to test algorithms for weakly chordal graphs. For instance, in~\cite{DBLP:journals/dmaa/MukhopadhyayRPG16} the authors generate all linear layouts of weakly chordal graphs. A generation mechanism can be used to obtain test instances for this algorithm. It can do the same for optimization algorithms, like finding a maximum clique, maximum stable set, minimum clique cover, minimum coloring, for both weighted and unweighted versions,
 for weakly chordal graphs propsed in~\cite{DBLP:journals/gc/HaywardHM89} and their improved versions in~\cite{DBLP:journals/talg/HaywardSS07,DBLP:journals/dam/SpinradS95}.\\
 
 If the input instances for a given algorithm are from a uniform distribution, a uniform random generation provides
 test instances to obtain an estimate of the average run-time of the algorithm. When the distribution is unknown, the
 assumption of uniform distribution might still help. Otherwise, we might look upon a generation algorithm as providing test-instances for an algorithm. With this motive, an algorithm for generating weakly chordal graphs by adding edges incrementally was recently proposed in \cite{DBLP:journals/corr/abs-1906-01056}. An application of this generation algorithm would be to obtain test-instances for an algorithm for enumerating linear layouts of a weakly chordal graph proposed in \cite{DBLP:journals/dmaa/MukhopadhyayRPG16}. \\
 
The next section of the paper contains some common graph terminology, used subsequently. The following section contains details of our algorithms, beginning with a brief overview. In the concluding section, we summarize the salient aspects of the paper and suggest directions for further work.

\section{Preliminaries}
We will assume that $G$ is a graph on $n$ vertices and $m$ edges, that is, $|V| = n$ and $|E| = m$. 
The {\em neighborhood} $N(v)$ of a vertex $v$ is the subset of vertices $\{u\in V\mid(u,v)\in E\}$ of $V$. The {\em degree} $\deg(v)$ of a vertex $v$ is equal to $|N(v)|$. A vertex $v$ of $G$ is {\em simplicial} if the induced subgraph on $N(v)$ is complete (alternately, a {\em clique}). A {\em path} in a graph $G$ is a sequence of vertices connected by edges. We use $P_k (k\geq 3)$ to denote a chordless path, spanning $k$ vertices of $G$. For instance, a path on 3 vertices is termed as a $P_3$ and, similarly, a path on 4 vertices is termed as a $P_4$. If a path starts and ends in the same vertex, the path is a cycle denoted by $C_k$, where $k$ is the length of the cycle. A {\em  chord} in a cycle is an edge between two non-consecutive vertices in the cycle. \\

\begin{figure}[htb]
	\centering
	\subfigure[{\em $G$} \label{Fig-CGExample}]{\includegraphics[scale=.65]{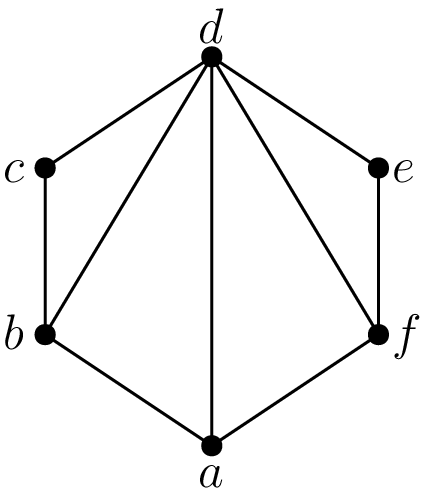}}\hspace{50pt}
	\subfigure[{\em $\overline{G}$} \label{Fig-CGCompExample}]{\includegraphics[scale=.65]{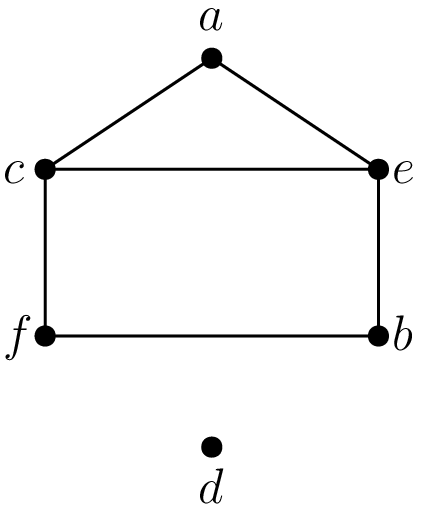}}
	\caption{{\em Complement of a chordal graph with a chordless 4-cycle~\cite{DBLP:journals/corr/abs-1906-01056}}}
	\label{Fig-CGExamples}
\end{figure}

\begin{figure}[h!]
	\centering
	\subfigure[{\em $G$} \label{Fig-FiveCycle}]{\includegraphics[scale=.61]{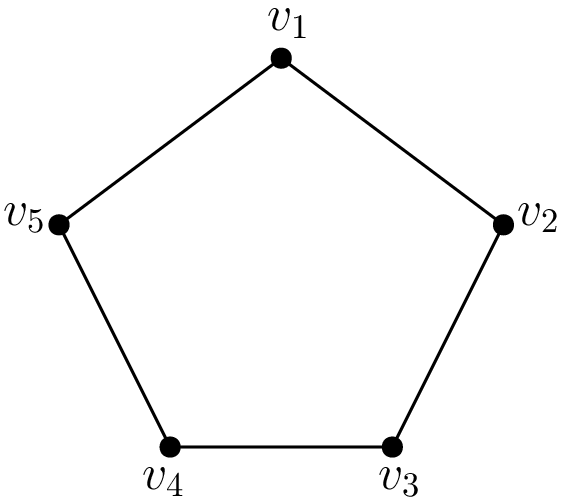}}\hspace{50pt}
	\subfigure[{\em $\overline{G}$} \label{Fig-FiveCycleComplement}]{\includegraphics[scale=.61]{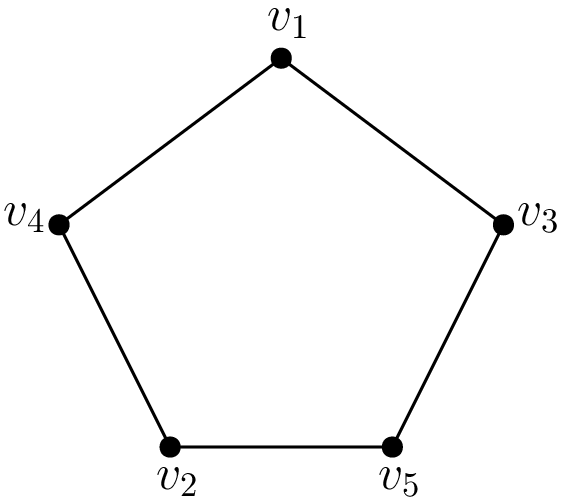}}
	\caption{{\em Complement of a five cycle is also a five cycle}}
	\label{Fig-FiveCycles}
\end{figure}

$G$ is chordal if it has no induced chordless cycles of size four or more. 
However, as Figure~\ref{Fig-CGExamples} shows, the complement of a chordal graph $G$ can contain an induced chordless cycle of size four. The complement cannot contain a five cycle though, as the complement of a five cycle is also a five cycle (see Figure~\ref{Fig-FiveCycles}). 
The above example makes it clear why chordal graphs are also weakly chordal.  \\

In this paper, we propose an algorithm that generates a weakly chordal graph from an arbitrary input graph. It is 
built on top of a subroutine that maintains the weak chordality of a graph $G$, under edge deletion. 


\section{Arbitrary Graph to Weakly Chordal Graph}\label{sec_arb_wcg}

\subsection{Overview of the Method}
\begin{figure}[htbp]
	\centering
	\includegraphics[scale=.45]{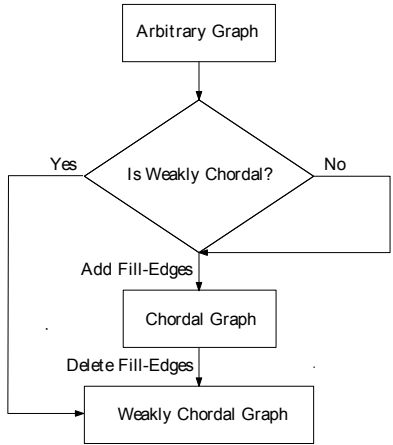}
	\caption{Overview of process}
	\label{Fig-Flowchart}
\end{figure}

We start by generating a random graph $G$ on $n$ vertices and $m$ edges. In a preprocessing step we check if $G$ is weakly chordal, 
using the LB-simpliciality recognition algorithm  due to \cite{DBLP:conf/swat/BerryBH00}. If $G$ is weakly chordal, we stop. Otherwise, we  proceed as follows. We first reduce $G$ to a chordal graph $H$ by introducing
additional edges, called fill-edges, using the minimum degree vertex ($mdv$, for short) heuristic \cite{DBLP:journals/siamrev/GeorgeL89}. The $mdv$ heuristic adds edges so that a minimum degree vertex in the current graph is simplicial.  Each fill-edge is also entered into a queue, termed a fill-edge queue, $FQ$. These fill-edges are potential candidates for subsequent deletion from $H$. Since $H$ is chordal, it is necessarily weakly chordal. We propose an algorithm for deleting edges from this weakly chordal graph to remove fill-edges, maintaining the weak chordality property. A fill-edge is deleted only if does not create a hole or an anti-hole in the resulting graph and we have developed criteria for detecting this.  A fill-edge is removed from the front of the queue, which we then try to delete. If we do not succeed we put it at the back of the queue. We keep doing this until no more fill edges can be removed. Operationally, we implement this by defining a deletion round as one in which the fill-edge at the back of the queue is at the front. We stop when the size of the queue does not change over two successive deletion rounds. 
Figure~\ref{Fig-Flowchart} is a pictorial illustration of the flow of control.

\subsection{Random Arbitrary Graph}
To generate a random graph, we invoke an algorithm by Keith M. Briggs, called `dense\_gnm\_random\_graph'. This algorithm,
based on  Knuth's Algorithm S (Selection sampling technique, see section 3.4.2 of~\cite{knuth1997}), takes the number of vertices, $n$ and the number of edges, $m$, as input and produces a random graph. For a given $n$, we set $m$ to a random value lying in the range between $n-1$ and $\frac{n(n-1)}{2}$. The output graph may be disconnected, in which case we connect the disjoint components, using additional edges.

\subsection{LB-simpliciality test}
In \cite{DBLP:conf/swat/BerryBH00} Berry et al. proved the following result: 

\begin{thm}\cite{DBLP:conf/swat/BerryBH00}
A graph is weakly chordal if and only if every edge is $LB$-simplicial. 
\end{thm}

We apply this recognition algorithm to the random graph generated by the previous step and continue with the next steps only if the recognition algorithm fails. Otherwise, we return $G$. 

\subsection{Arbitrary Graph to Chordal Graph} 
The arbitrary graph $G$ is embedded into a chordal graph $H$ by the addition of edges and the process is known as triangulation or fill-in.
Desirable triangulations are those in which a minimal or a minimum number of edges is added. 
  A triangulation $H = (V, E \cup F) $ of $G = (V,E)$ is minimal if $(V, E \cup F')$ is non-chordal for every proper subset $F'$ of $F$. In a minimum triangulation the number of edges added is the fewest possible. Berry at al. \cite{DBLP:conf/soda/Berry99}  proposed an algorithm, known as LB-Triangulation, for the minimal fill-in problem.  LB-Triangulation works on any ordering $\alpha$ of the vertices, and produces a fill that is provably exclusion-minimal. In our algorithm, we have used the {\em mdv} heuristic~\cite{DBLP:journals/siamrev/GeorgeL89}, as our experiments have shown that 
  this adds fewer fill-edges as compared to LB-Triangulation. We explain this heuristic in the next section. 

\subsubsection{The Minimum Degree Vertex Heuristic} 
Let $H = (V, E \cup F)$ be the graph obtained from $G = (V, E)$, where $F$ is set of fill-edges, folllowing these steps. We first assign $G$ to $H$ and then
prune from $G$ all vertices of degree 1.
From the remaining vertices of $G$ we choose a vertex $v$ of minimum degree (breaking ties arbitrarily) and turn the neighborhood $N(v)$ of $v$ into a clique by adding edges. These are fill-edges that we add to the edge set of $H$, as well as to the fill-queue, $FQ$. Finally, 
we remove from $G$, the vertex $v$ and all the edges incident on it. We repeat this until $G$ is empty. The graph $H$ is now chordal and is identical with the initial graph $G$, sans degree 1 vertices, and with fill edges added. We illustrate this with an example. 

\begin{figure}[htb]
	\centering
	\subfigure[{\em Arbitrary graph} \label{Fig-MDPic1}]{\includegraphics[scale=.65]{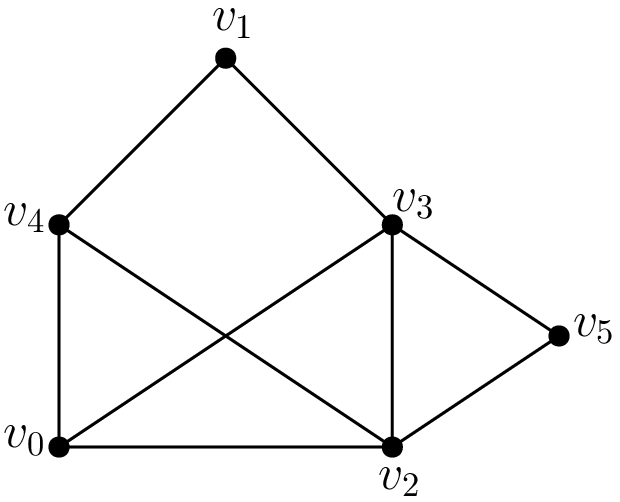}}\hspace{50pt}
	\subfigure[{\em Chordal Graph} \label{Fig-MDPic2}]{\includegraphics[scale=.65]{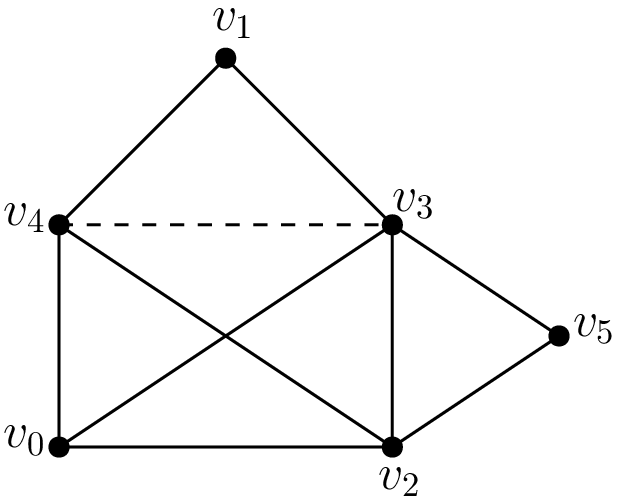}}
	\caption{{\em Arbitrary graph to chordal graph}}
	\label{Fig-MDExamples}
\end{figure}

The initial graph $G$ is shown in Fig.~\ref{Fig-MDPic1} and the graph $H$ with all fill-edges added is shown in Fig.~\ref{Fig-MDPic2}. In the initial graph $G$ both $v_1$ and $v_5$ have minimum degree. We break tie in favour of $v_5$. 
Since the induced subgraph on $N(v_5)$ is already a clique no fill-edges are added and $G$ is set to $G -\{v_5\}$. In the reduced 
graph $G$, $v_1$ is of minimum degree and the induced graph on $N(v_1)$ is turned into a clique by adding $\{v_3, v_4\}$ as a fill- edge, which is also added to $H$. Since the reduced graph $G - \{v_1\}$ is a clique, we can pick the vertices $v_0, v_2, v_3, v_4$
in an arbitrary order to reduce $G$ to an empty graph, without introducing any further fill edges into $H$. The formal algorithm is described below: \\ 

\begin{algorithm}[htb]
	\caption{ArbitraryToChordal}\label{algoArbToCG}
	\begin{algorithmic}[1]
		\Require An arbitrary graph $G=(V,E)$
		\Ensure Returns a chordal graph $H=(V,E\cup F)$ and fill-edge queue $FQ$
		\State $H \leftarrow G$
		\State Delete all vertices of degree 1 from $G$
		\State Sort $V$ in ascending order of degrees \label{sort}
		\State Choose a vertex $v$ of minimum degree \label{choose}
		\State Turn $N(v)$ of $v$ into a clique by adding edges, which are added to the edge set of $H$ and to the fill-queue, $FQ$ \label{turn} 
		\State Remove the vertex $v$ from $G$ and all the edges incident on it \label{remove}
		\State Repeat steps~\ref{sort} to~\ref{remove} until $G$ is empty
	\end{algorithmic}
\end{algorithm}

\subsection{Chordal Graph to Weakly Chordal Graph}

Since the chordal graph $H$ obtained from the previous stage is also weakly chordal, we apply an edge deletion algorithm 
to $H$ that preserves weak chordality. The edges that are candidates for deletion are the ones that have been added by the $mdv$ heuristic. Each candidate edge is temporarily deleted from $H$, and we check if its deletion creates a hole or an anti-hole in $H$. If not, we  delete this edge. The process is explained in details in the subsequent sections. 

\subsubsection{Fill-Edge Queue}
As mentioned earlier, each edge added to convert an arbitrary input graph into chordal graph is called a fill-edge. In order to delete as many fill-edges as possible, we maintain a queue of fill-edges, $FQ$. A fill-edge is removed from the front of this queue, which we then try to delete from $H$. If we do not succeed because a hole or anti-hole is created, we put it at the back of the queue. We keep doing this until no more fill-edges can be removed from $FQ$.

\begin{figure}[htb]
	\centering
	\subfigure[{\em One $P_4$ and one $P_3$} \label{Fig-Hole34}]{\includegraphics[scale=.65]{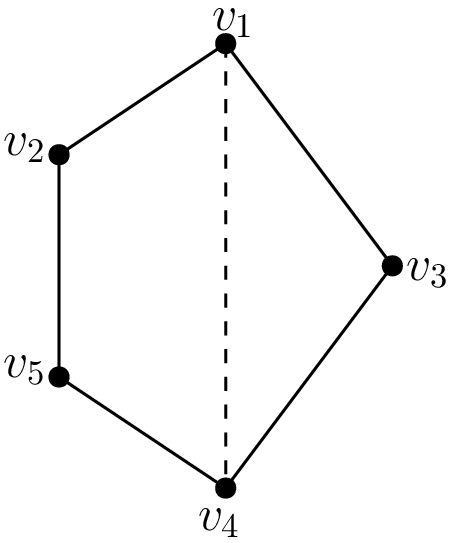}}\hspace{50pt}
	\subfigure[{\em Two $P_4$} \label{Fig-Hole44}]{\includegraphics[scale=.65]{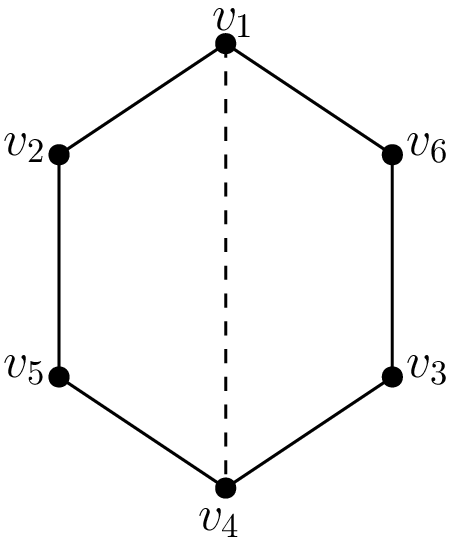}}
	\caption{{\em Detecting Holes}}
	\label{Fig-DetectingHoles}
\end{figure}

\subsubsection{Detecting Holes}

To reiterate, a hole in a graph $G$ is an induced chordless cycle on five or more vertices. Since, a graph is weakly chordal if it is (hole, anti-hole)-free~\cite{DBLP:journals/corr/abs-1902-08071}, it is crucial to detect if any hole is formed by the deletion of an edge. For the class of weakly chordal graphs, since the biggest cycle allowed is of size four, the holes can be formed either by a combination of two $P_4$'s or a by a combination of a $P_3$ and a $P_4$, as illustrated in  Fig.~\ref{Fig-DetectingHoles}. \\

To detect the formation of a hole in $H$, we pick an edge $e=\{u,v\}$ of $H$ and temporarily delete it. Now, we check if this deletion  creates a hole in $H$. To detect a hole, we perform a breadth-first search in $H$ with $u$ as the source vertex and find all chordless $P_3$ and $P_4$ paths between $u$ and $v$. A hole can be created in two distinct ways: (i) by a disjoint pair of $P_4$, with six distinct vertices between them such that there exist no chord joining an internal vertex on one $P_4$ to an internal vertex on the other; this we call  a hole on two $P_4$s; (ii) by a disjoint pair of $P_3$ and $P_4$ between $u$ and $v$, with five distinct vertices between them, such that there exist no chord joining an internal vertex on the $P_4$ to the internal vertex of the $P_3$; this we call a hole on a $P_3$ and a $P_4$. 

\subsubsection{Antiholes}
\begin{figure}[h!]
	\centering
	\includegraphics[scale=.63]{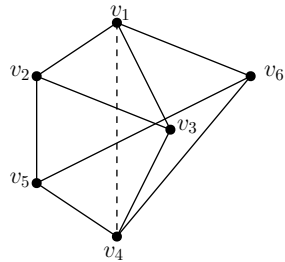}
	\caption{Antihole}
	\label{Fig-AntiHole}
\end{figure}

An anti-hole in a graph is, by definition, the complement of a hole~\cite{DBLP:journals/corr/abs-1902-08071}. An anti-hole configuration in a weakly chordal graphs has the structure shown in Fig.~\ref{Fig-AntiHole}. This is an induced graph on  six distinct vertices each of which is of degree three. 


\subsubsection{Detecting Antiholes}
To detect an anti-hole configuration, we pick an edge $\{u,v\}$ and temporarily delete it from the graph. Next, we check if deleting the edge $\{u,v\}$ creates an anti-hole configuration in the graph. To detect this, we do breadth-first search with $u$ as the 
source vertex to find all chordless $P_3$ and $P_4$ paths between $u$ and $v$. An anti-hole configuration is formed by a combination of two $P_3$ and one $P_4$ such that the induced graph on the six vertices that define these paths, are uniformly of degree three and there exists a chord from the internal vertex of each $P_3$ to one of the internal vertices in the $P_4$. For example, in  Fig.~\ref{Fig-AntiHole}, $\{v_1,v_2,v_5,v_4\}$ is a $P_4$, $\{v_1,v_3,v_4\}$ and $\{v_1,v_6,v_4\}$ are two $P_3$ paths. There exists exactly one chord from $v_2$ to $v_3$ and exactly one from $v_5$ to $v_6$ and, in the induced graph on these six vertices, every vertex has degree three, making it an anti-hole configuration.

\subsubsection{Proposed Algorithm}
We use an algorithm for deleting edges from a weakly chord graph to remove fill edges, maintaining its weak chordality property. In order to delete as many fill-edges as possible,  a fill-edge $\{u, v\}$  is removed from the front of the fill-queue, which we then try to delete from $H$. If we do not succeed, we put it at the back of the queue. We keep doing this until no more fill-edges can be removed. Operationally, we implement this by defining a deletion round as one in which the edge at the back of the queue is at the front.
One deletion round comprises of picking an edge from the start of the queue and deleting it from $H$. Now we check if the deletion of  $\{u, v\}$ creates a hole or an antihole in $H$. If so, we do not delete the edge 
 $\{u, v\}$ and add it back to the fill-queue. Otherwise,  we delete the edge from $H$ and also remove it from te fill-queue $FQ$. We stop when the size of $FQ$ does not change over two successive deletion rounds. \\

%
%
%
%
%

\begin{algorithm}[htb]
	\caption{ChordalToWeaklyChordal}\label{algoCGToWCG}
	\begin{algorithmic}[1]
		\Require A chordal graph $H=(V,E\cup F)$ with fill edge queue $F$
		\Ensure A weakly chordal graph $G_w$
		\State $T \leftarrow H$ \Comment{Make a copy of $H$}
		\State $FQ \leftarrow$ fill-edges of $H$ 
		\State $prevSize \leftarrow  0$ 
		\State $newSize \leftarrow  |FQ|$ 
		\While {($prevSize \neq newSize$ \&\& $newSize \neq 0$)} \Comment{Check size of $FQ$ over two deletion rounds}
			\State $prevSize \leftarrow newSize$
			\For{($each ~edge \{u,v\} ~ in fill$-$queue$, $FQ$)}
				\State Delete edge $\{u,v\}$ from $T$
				\If {($Hole ~or~ Antihole ~ Detected$)}
					\State Do not delete edge from graph $H$, add edge back to temporary graph $T$, 
						and to the back of the queue $FQ$ 
				\Else
					\State Delete edge $\{u,v\}$ from graph $H$ 
				\EndIf
			\EndFor
			\State $newSize \leftarrow |FQ|$
		\EndWhile
		\State $G_w \leftarrow H$
		\State {\bf return} $G_w$
	\end{algorithmic}
\end{algorithm}

\begin{figure}[htb]
	\centering
	\subfigure[{\em Arbitrary graph ($G$)} \label{Fig-HoleExample1}]{\includegraphics[scale=.65]{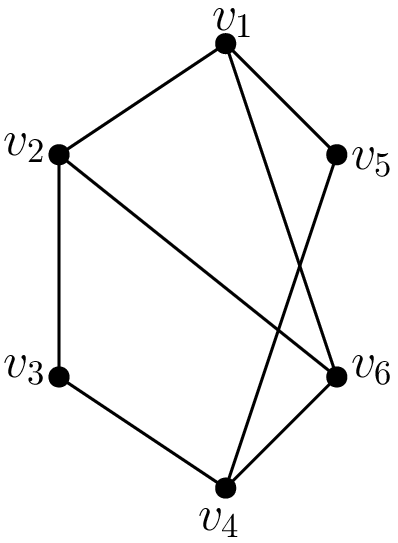}}\hspace{50pt}
	\subfigure[{\em Chordal graph ($H$)} \label{Fig-HoleExample2}]{\includegraphics[scale=.65]{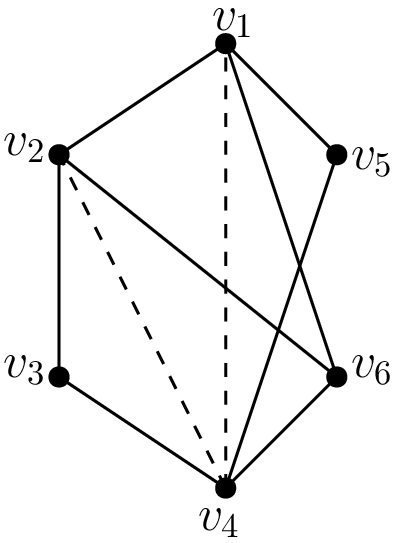}}\hspace{50pt}
	\subfigure[{\em Weakly chordal graph ($H = H-\{v_2,v_4\}=G_w$)} \label{Fig-HoleExample3}]{\includegraphics[scale=.65]{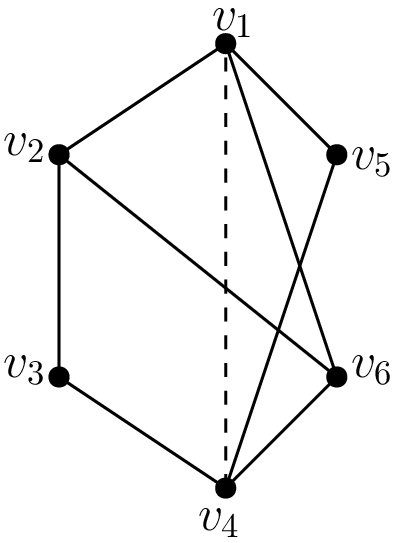}}
	\caption{{\em Arbitrary graph to weakly chordal graph}}
	\label{Fig-ArbToWCGHoles}
\end{figure}

\vspace{0.2cm}
For example in Fig.~\ref{Fig-ArbToWCGHoles}, a random arbitrary graph on 6 vertices and 8 edges is obtained. It is converted into a chordal graph by inserting two additional edges. These two additional edges added are put in the fill-edge queue $[\{v_2,v_4\},\{v_1,v_4\}]$. Maintain a temporary copy of chordal graph $G$ in $T$. The deletion algorithm begins by picking first edge $\{v_2,v_4\}$ in the fill-edge queue and temporarily deletes it from graph $T$ to check for hole and antihole configurations. Since deleting $\{v_2,v_4\}$ does not give rise to any hole or anti-hole configurations, $\{v_2,v_4\}$ is permanently deleted from starting graph $H$ which is now a weakly chordal graph. Now the updated fill-edge queue is $[\{v_1,v_4\}]$. The deletion algorithm now picks the first edge in $\{v_1,v_4\}$ in the fill edge queue and temporarily deletes it from graph $T$ to check for hole and antihole configurations. Since deleting $\{v_1,v_4\}$ gives rise to a hole configuration on one $P_4$ $\{v_1,v_2,v_3,v_4\}$ and one $P_3$ $\{v_1,v_5,v_4\}$, $\{v_1,v_4\}$ is not permanently deleted from $H$. Since the queue is now empty, the graph $G_w$ returned by the algorithm is weakly chordal with a small subset of fill-edges added to the original graph $G$. \\

\begin{figure}[h!]
	\centering
	\subfigure[{\em Arbitrary graph ($G$)} \label{Fig-AntiholeExample1}]{\includegraphics[scale=.59]{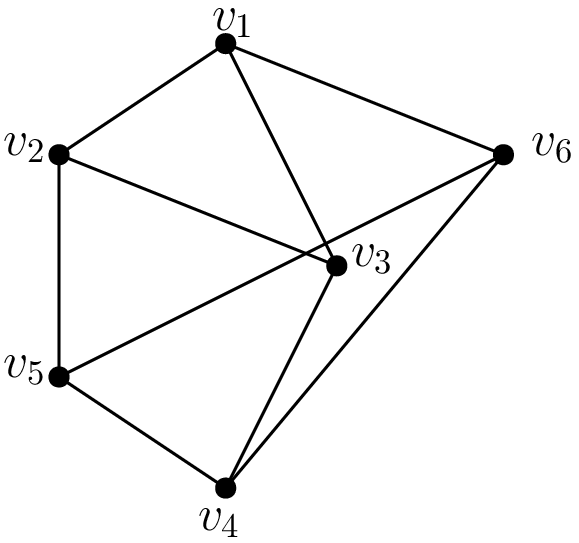}}\hspace{50pt}
	\subfigure[{\em Chordal graph ($H$)} \label{Fig-AntiholeExample2}]{\includegraphics[scale=.59]{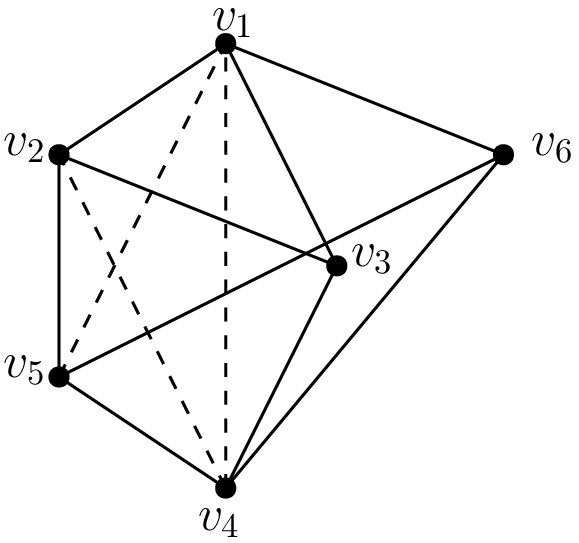}}\\
	\subfigure[{\em Chordal graph ($H = H-\{v_1,v_5\}$)} \label{Fig-AntiholeExample3}]{\includegraphics[scale=.59]{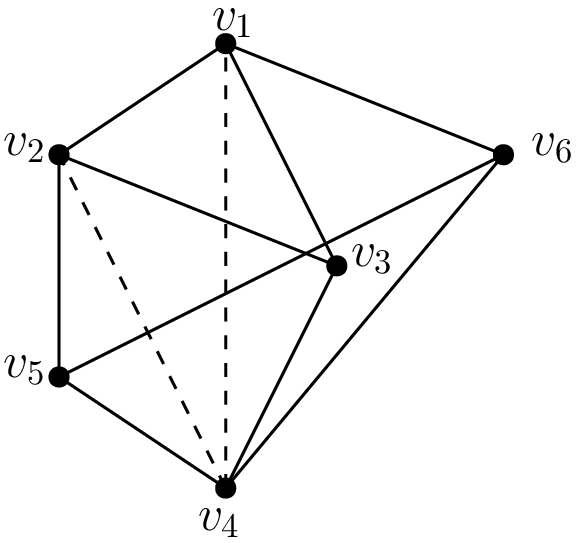}}\hspace{50pt}
	\subfigure[{\em Weakly chordal graph ($H = H-\{v_2,v_4\}=G_w$)} \label{Fig-AntiholeExample4}]{\includegraphics[scale=.59]{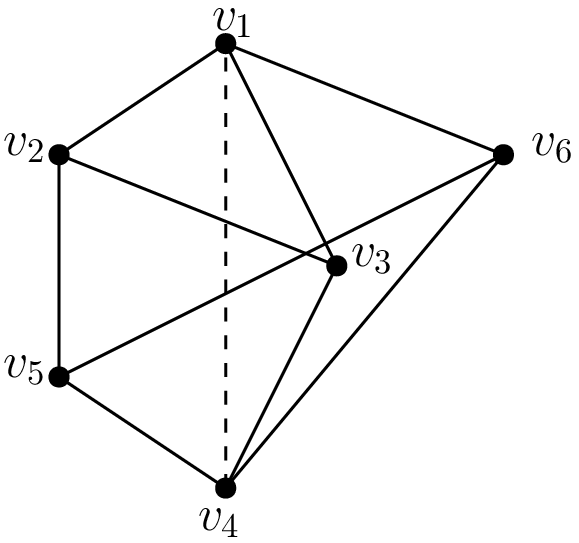}}
	\caption{{\em Arbitrary graph to weakly chordal graph}}
	\label{Fig-ArbToWCGAntiholes}
\end{figure}

For another example, consider Figure~\ref{Fig-AntiholeExample1}, a random arbitrary graph on 6 vertices and 9 edges is obtained. It is converted into a chordal graph $H$ (see Figure~\ref{Fig-AntiholeExample2}) by adding three additional edges. These three additional edges added are put in the fill-edge queue $[\{v_1,v_5\},\{v_2,v_4\},\{v_1,v_4\}]$. Maintain a temporary copy of the chordal graph $H$ in $T$. The deletion algorithm begins by picking first edge $\{v_1,v_5\}$ in the fill edge queue and temporarily deletes it from graph $T$ to check for hole and antihole configurations. Since deleting $\{v_1,v_5\}$ does not give rise to any hole or anti-hole configurations, $\{v_1,v_5\}$ is permanently deleted from starting graph $H$ which is now a weakly chordal graph shown in Figure~\ref{Fig-AntiholeExample3}. Now the updated fill-edge queue is $[\{v_2,v_4\},\{v_1,v_4\}]$. The deletion algorithm now picks the first edge in $\{v_2,v_4\}$ in the fill-edge queue and temporarily deletes it from graph $T$ to check for hole and anti-hole configurations. Since deleting $\{v_2,v_4\}$ does not give rise to any hole or anti-hole configuration, $\{v_2,v_4\}$ is permanently deleted from starting graph $H$, which is now a weakly chordal graph. Now the updated fill-edge queue is [$\{v_1,v_4\}$]. The deletion algorithm now picks the first and only edge $\{v_1,v_4\}$ in the fill-edge queue and temporarily deletes it from graph $T$ to check for a hole or an anti-hole configuration. Since deleting $\{v_1,v_4\}$ gives rise to an anti-hole configuration on two $P_3$ paths $\{v_1,v_3,v_4\}$,$\{v_1,v_6,v_4\}$ and one $P_4$ $\{v_1,v_2,v_5,v_4\}$, the edge $\{v_1,v_4\}$ is not permanently deleted from starting graph $H$. Since the queue is now empty, the graph $G_w$ returned by the algorithm is weakly chordal with a small subset of fill-edges added to the original graph $G$ as shown in Figure~\ref{Fig-AntiholeExample4}.

\subsection{Complexity}
The $mdv$ heuristic can be implemented in $O(n^2 m)$ time, while the time-complexity of the recognition algorithm based 
on LB-simpliciality is in $O(nm)$. \\

\begin{figure}[htb]
	\centering
	\includegraphics[scale = .65]{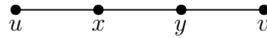}
	\caption{{\em A $P_4$-path from $u$ to $v$}}
	\label{Fig-AP4Path}
\end{figure}

To bound the query complexity of deleting an edge $\{u, v\}$ from the weakly chordal graph, we note that this is dominated by the task of finding multiple $P_3$ and $P_4$ paths between $u$ and $v$ and we have to consider these in pairs and run the breadth-first search. An upper bound on the number of pairs of $P_3$ and $P_4$ paths between $u$ and $v$ is $O(d_u^2d_v^2)$, where $d_u$ and $d_v$ are the degrees of $u$ and $v$ respectively. For consider such a path from $u$ to $v$ (see Figure~\ref{Fig-AP4Path}): $x$ is one of the at most $d_u$ vertices adjacent to $u$ and $y$ is one of the at most $d_v$ vertices adjacent to $v$, so that we have at most $O(d_ud_v)$ $P_4$ paths from $u$ to $v$ and thus $O(d_u^2d_v^2)$ disjoint pairs of $P_4$ paths from $u$ to $v$.\\

If $|E|$ be the number of edges currently, in the weakly chordal graph, the complexity of running a breadth-first search is $O(n + |E|)$. Since $m$ is the number of edges in the final weakly chordal graph, an upper bound on the query complexity is $O(d_u^2d_v^2 (n + m))$. \\

The deletion of an edge take constant time since we maintain an adjacency matrix data structure to represent $G$.

\section{Conclusion}
We have proposed a simple method for generating a weakly chordal graph from an arbitrary graph. The proposed algorithm can also be used to generate weakly chordal graphs by deleting edges from input graphs that are known to be weakly chordal, such as complete graphs. Starting with complete graphs also helps in generating dense weakly chordal graphs. An interesting open problem is to establish if the proposed method to generate a weakly chordal graph from an arbitrary graph adds a minimal number of edges. \\

We have implemented our algorithm in Python. Some sample outputs are shown below in an appendix. In each of the figures, 
the purples edges in the chordal graph are edges that are candidates for deletion.\\


\newpage
\section{Appendix}
\begin{figure}[htb]
	\centering
	\subfigure[{\em $Arbitrary~Graph$} \label{sample1.1}]{\includegraphics[scale=.35]{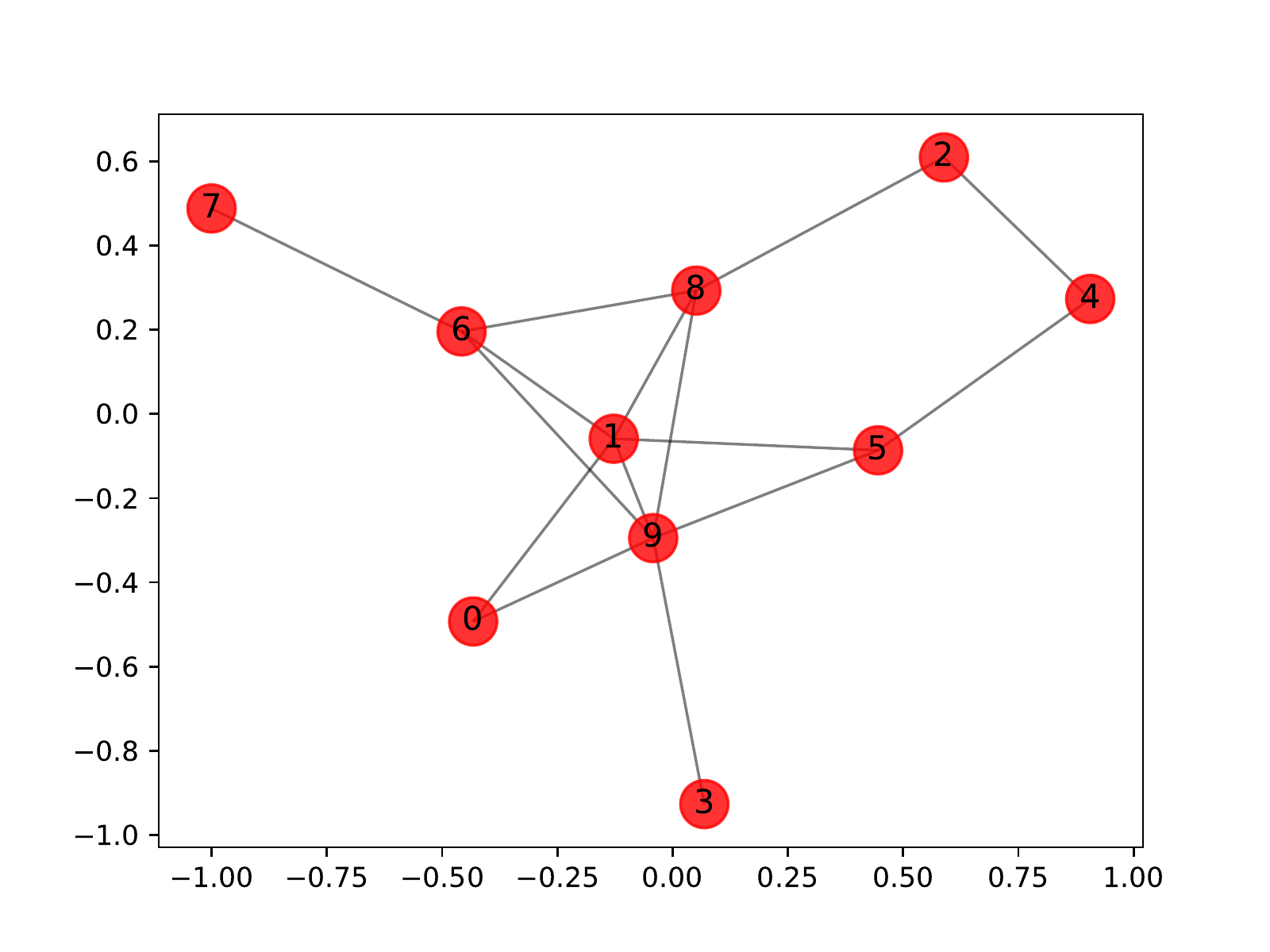}}\hspace{50pt}
	\subfigure[{\em $Chordal~Graph$} \label{sample1.2}]{\includegraphics[scale=.35]{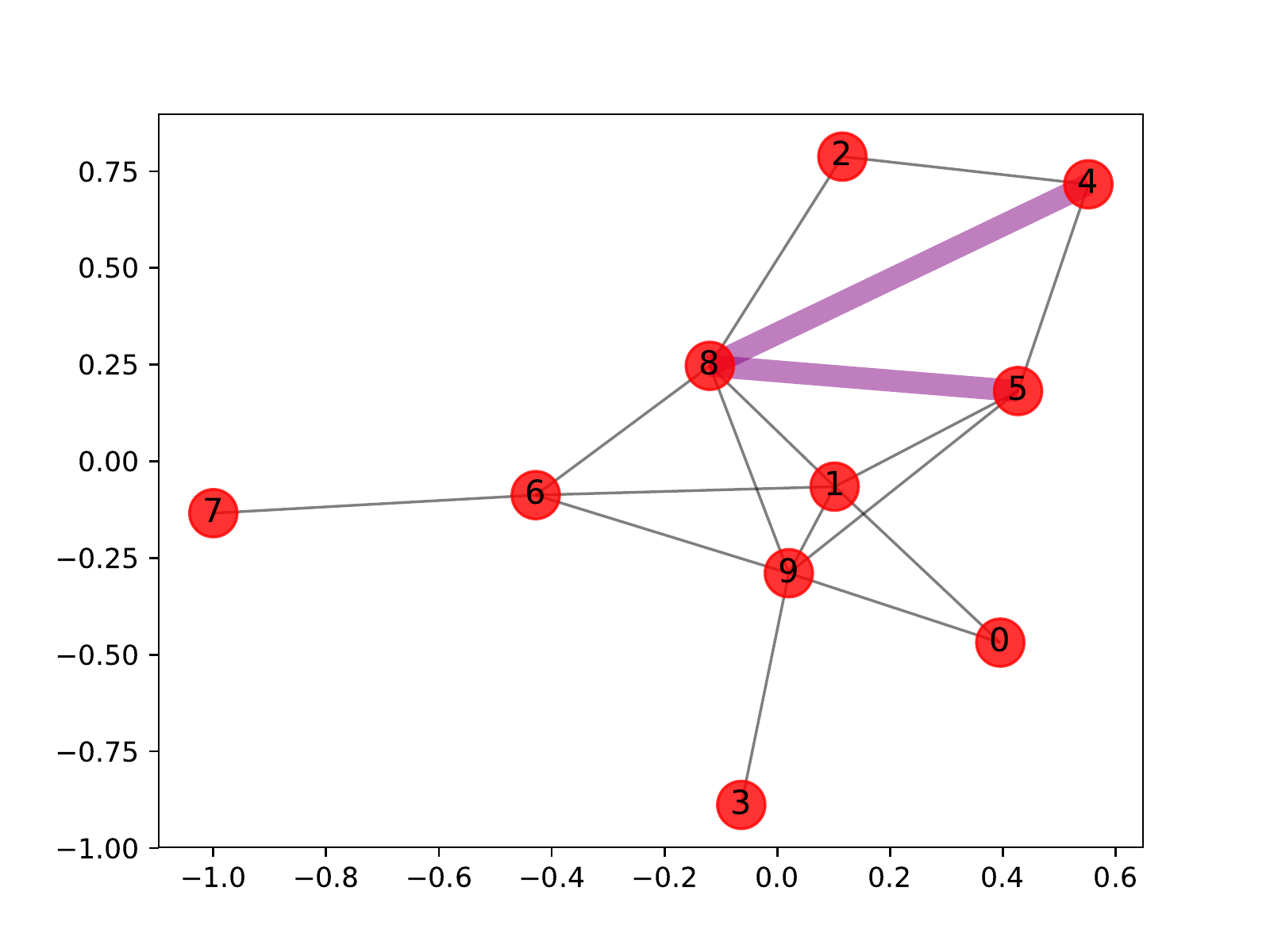}}\hspace{50pt}
	\subfigure[{\em $Weakly~Chordal~Graph$} \label{sample1.3}]{\includegraphics[scale=.35]{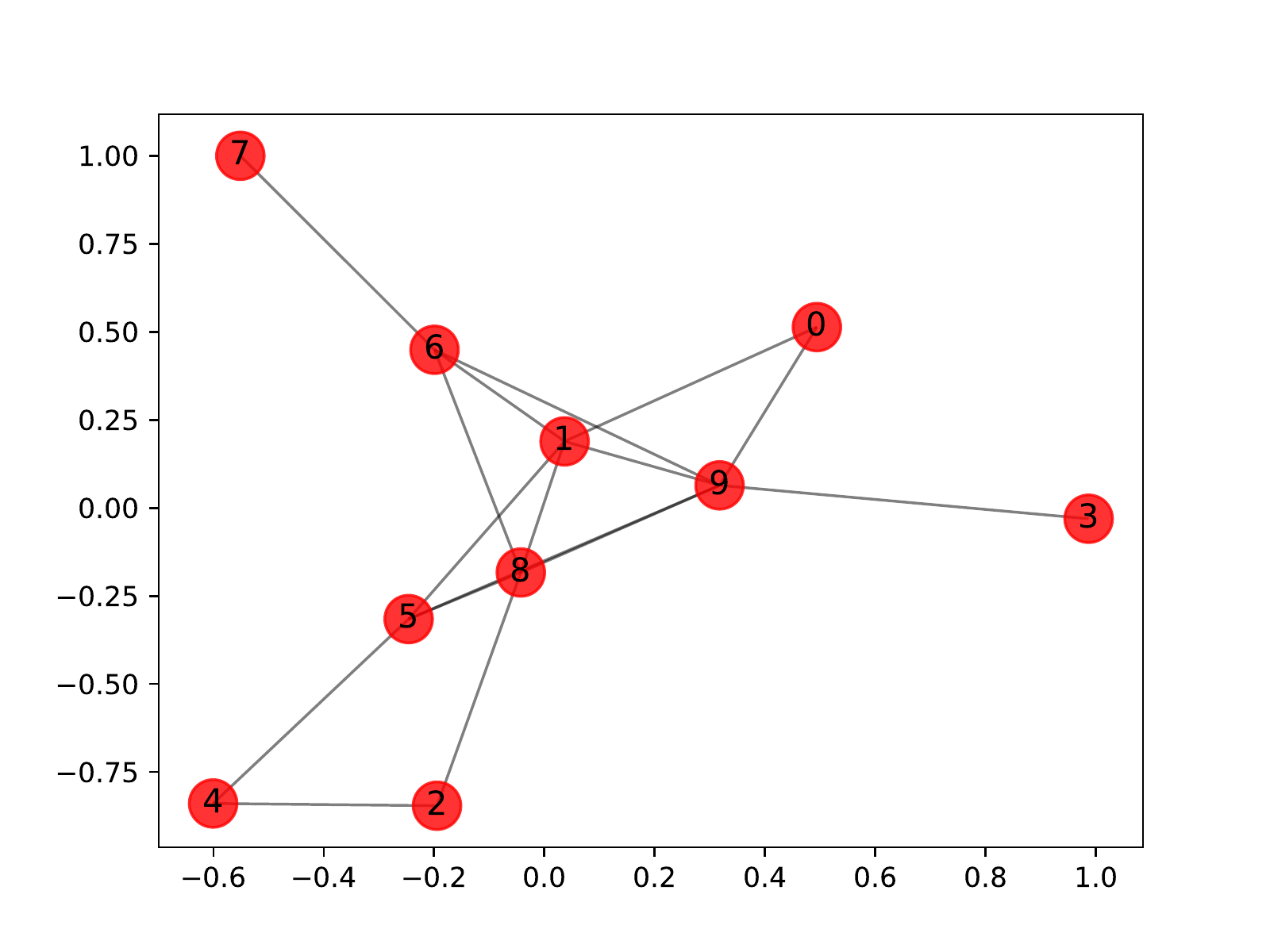}}\hspace{50pt}
	\caption{{\em Arbitary graph to a weakly chordal one}}
	\label{Sample1}
\end{figure}

\begin{figure}[htb]
	\centering
	\subfigure[{\em $Arbitrary~Graph$} \label{sample2.1}]{\includegraphics[scale=.35]{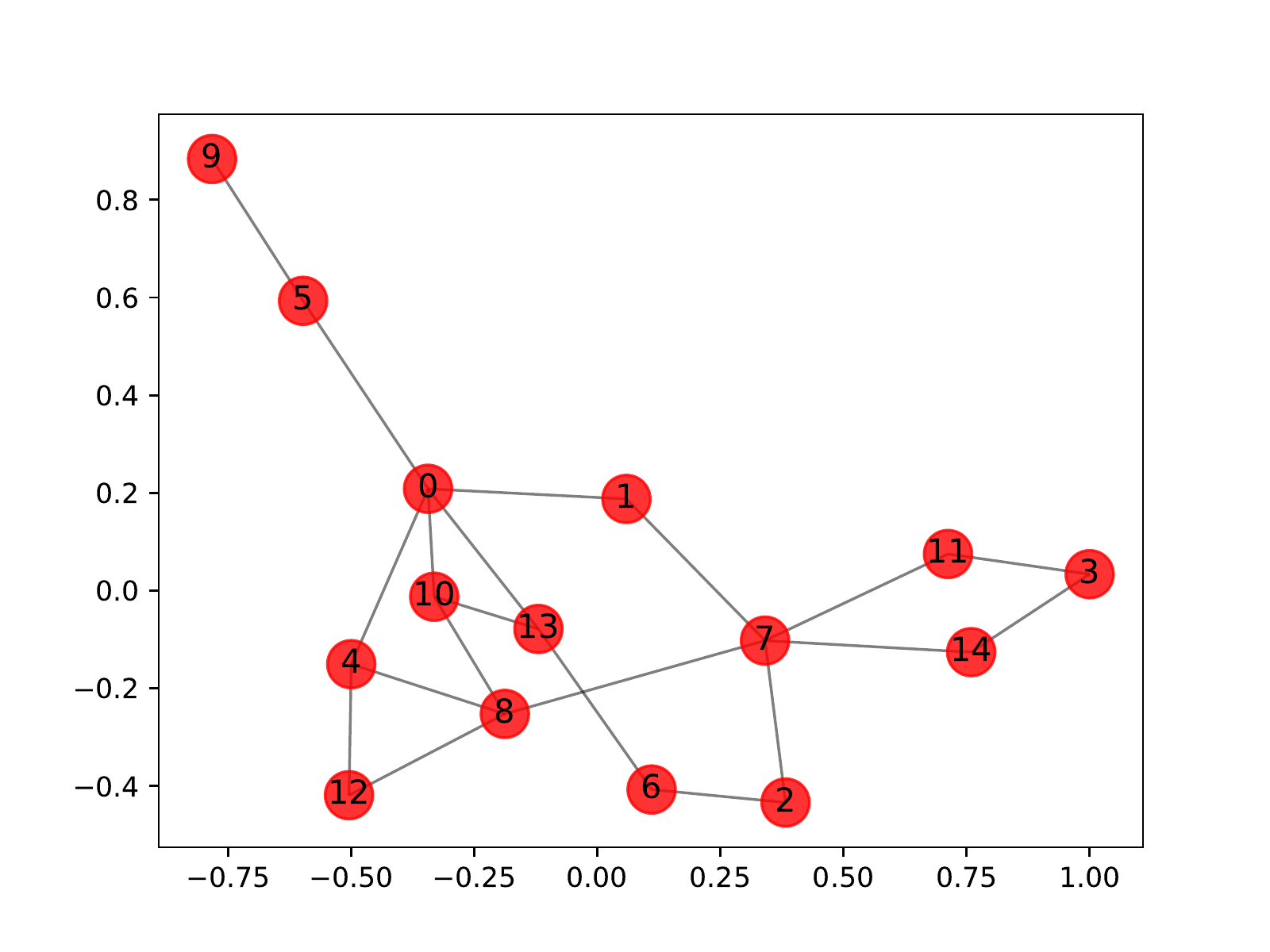}}\hspace{50pt}
	\subfigure[{\em $Chordal~Graph$} \label{sample2.2}]{\includegraphics[scale=.35]{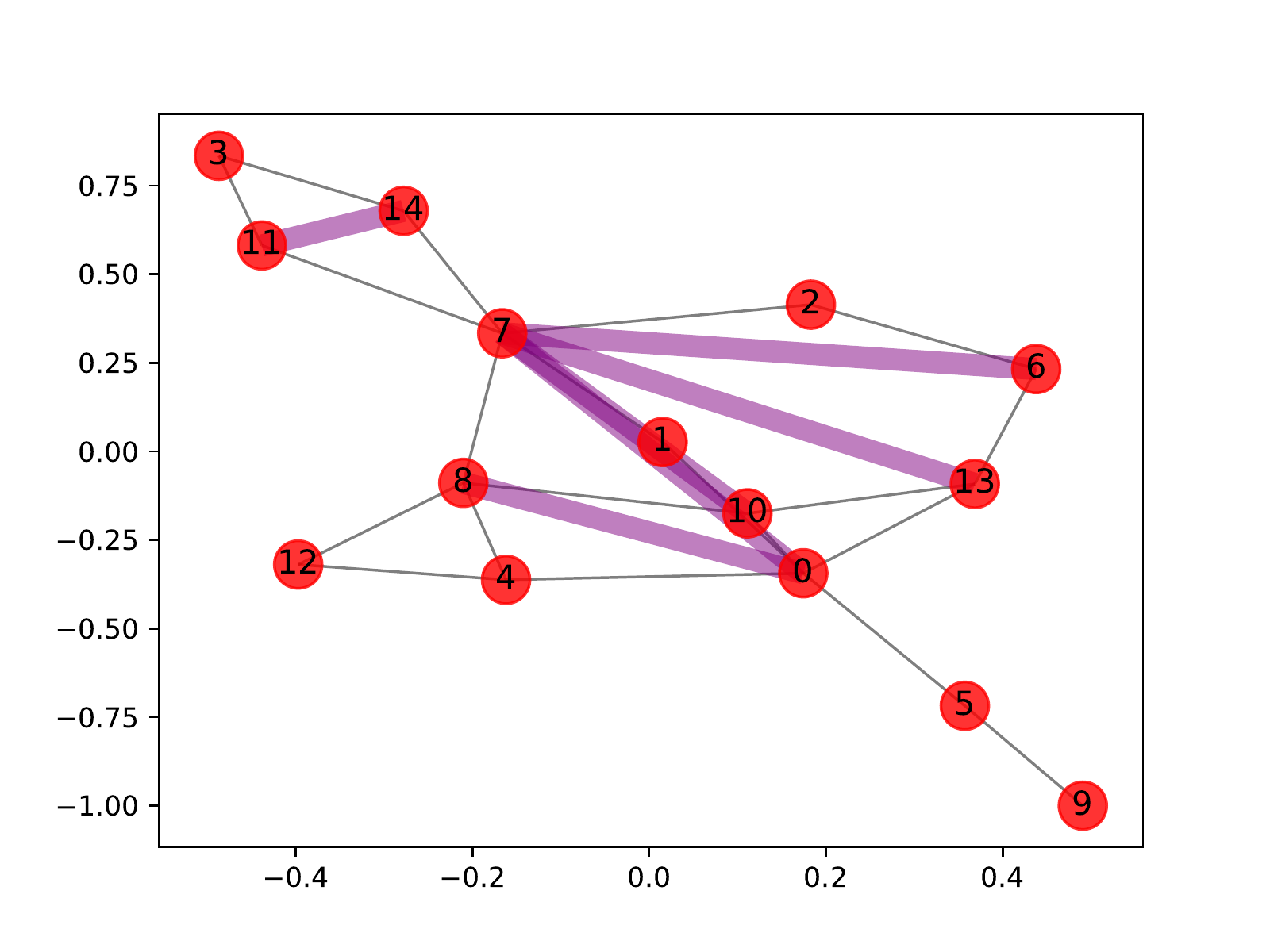}}\hspace{50pt}
	\subfigure[{\em $Weakly~Chordal~Graph$} \label{sample2.3}]{\includegraphics[scale=.35]{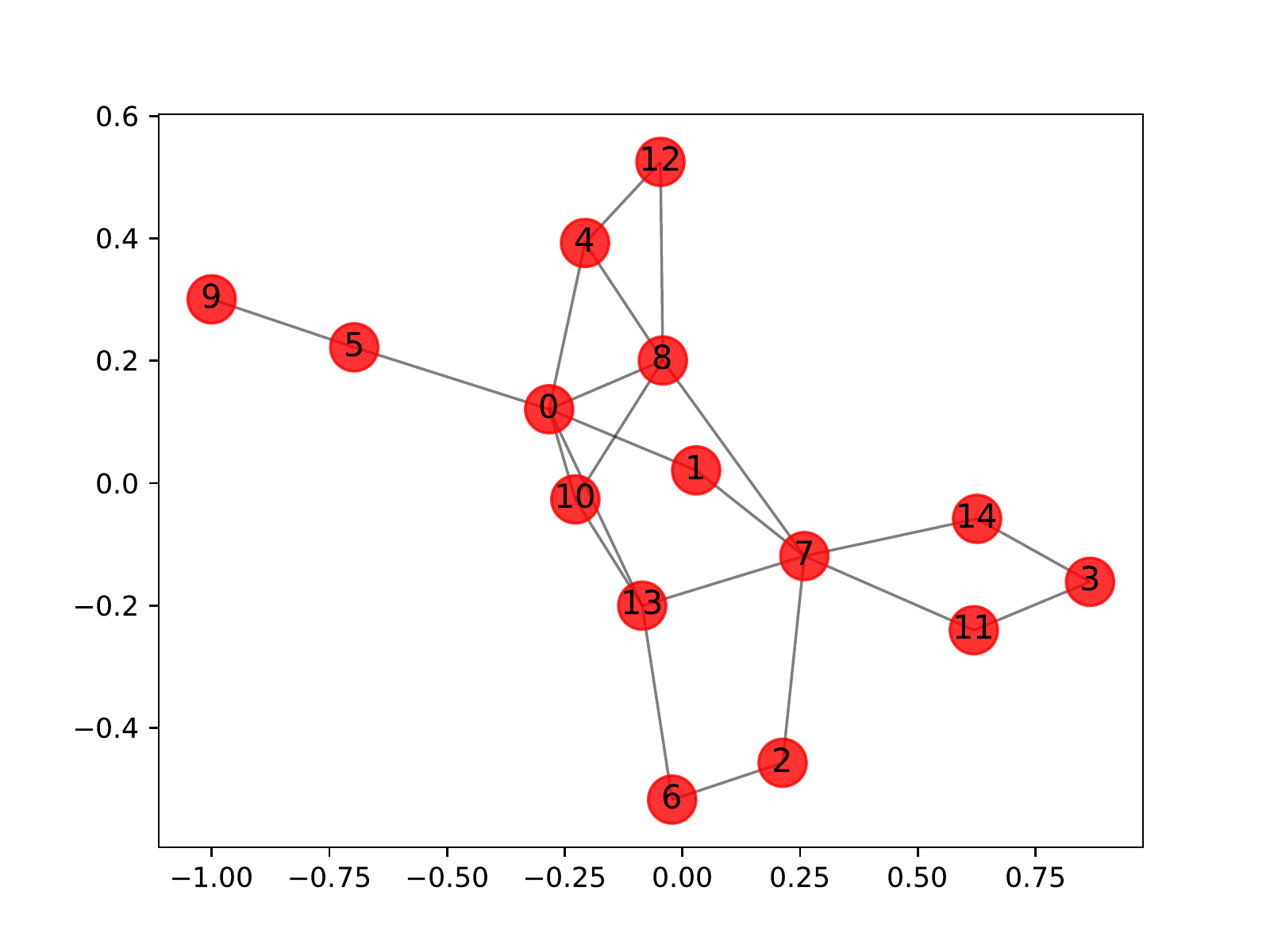}}\hspace{50pt}
	\caption{{\em Arbitary graph to a weakly chordal one}}
	\label{Sample2}
\end{figure}
\begin{figure}[htb]
	\centering
	\subfigure[{\em $Arbitrary~Graph$} \label{sample3.1}]{\includegraphics[scale=.35]{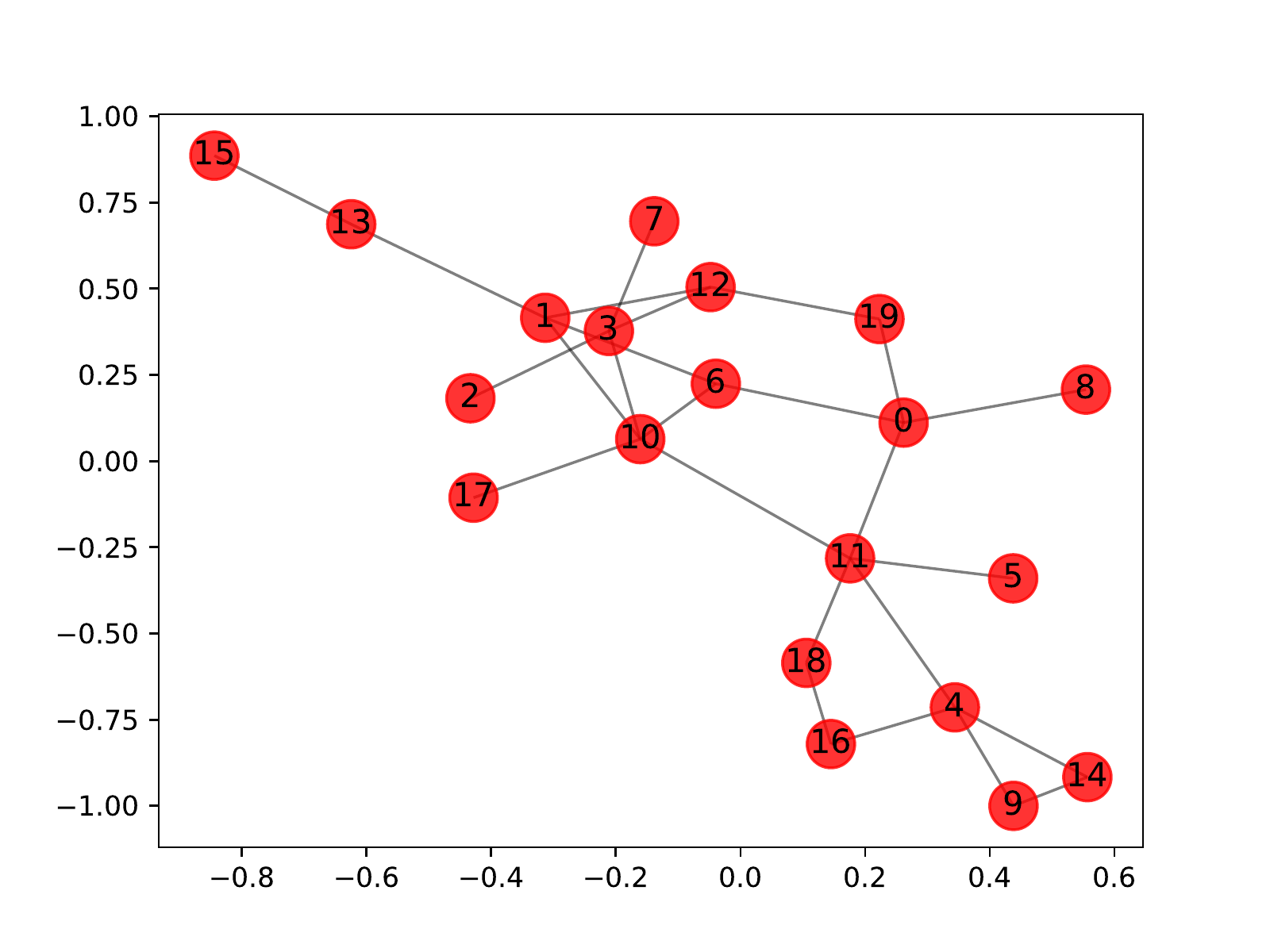}}\hspace{50pt}
	\subfigure[{\em $Chordal~Graph$} \label{sample3.2}]{\includegraphics[scale=.35]{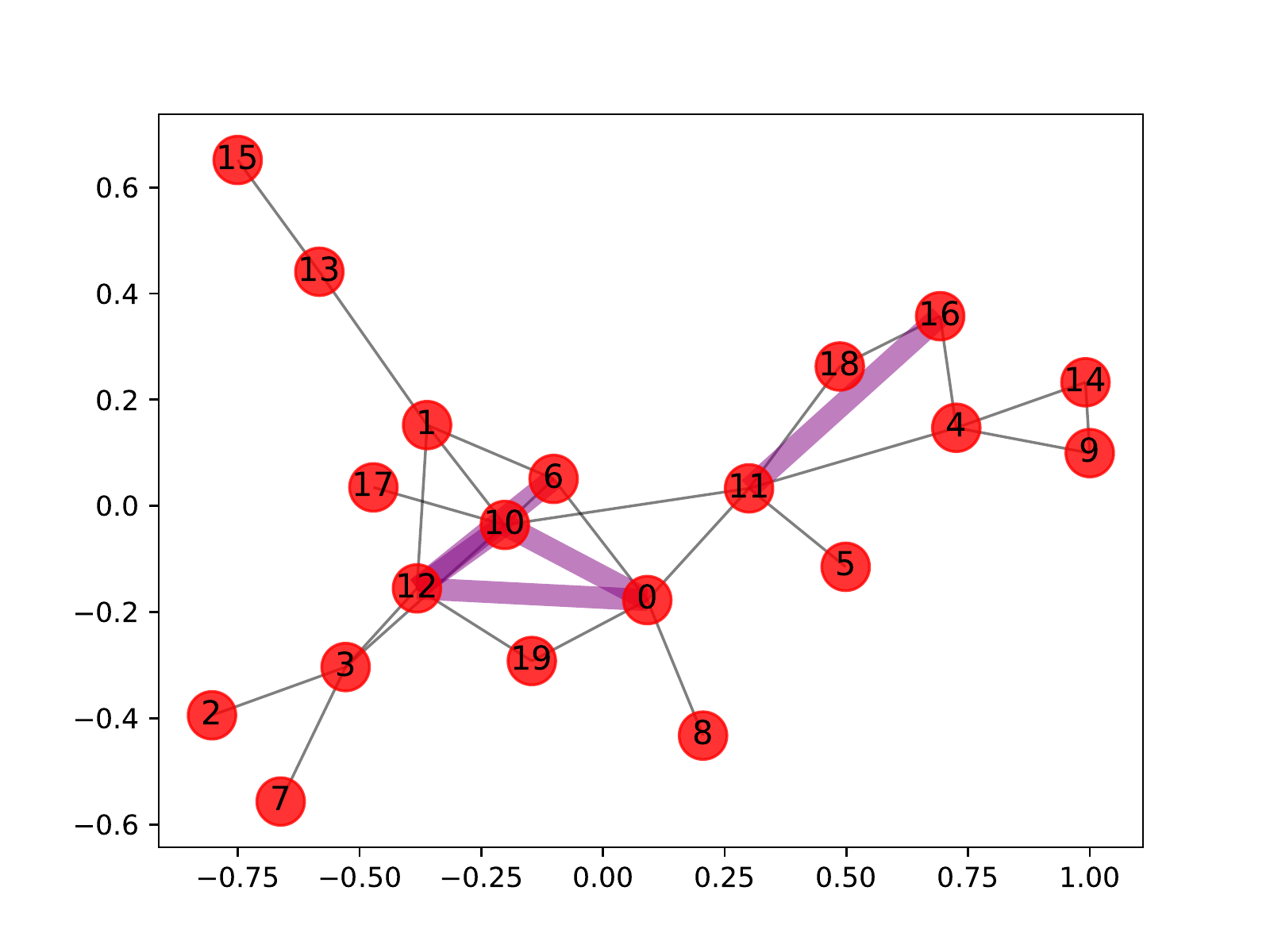}}\hspace{50pt}
	\subfigure[{\em $Weakly~Chordal~Graph$} \label{sample3.3}]{\includegraphics[scale=.35]{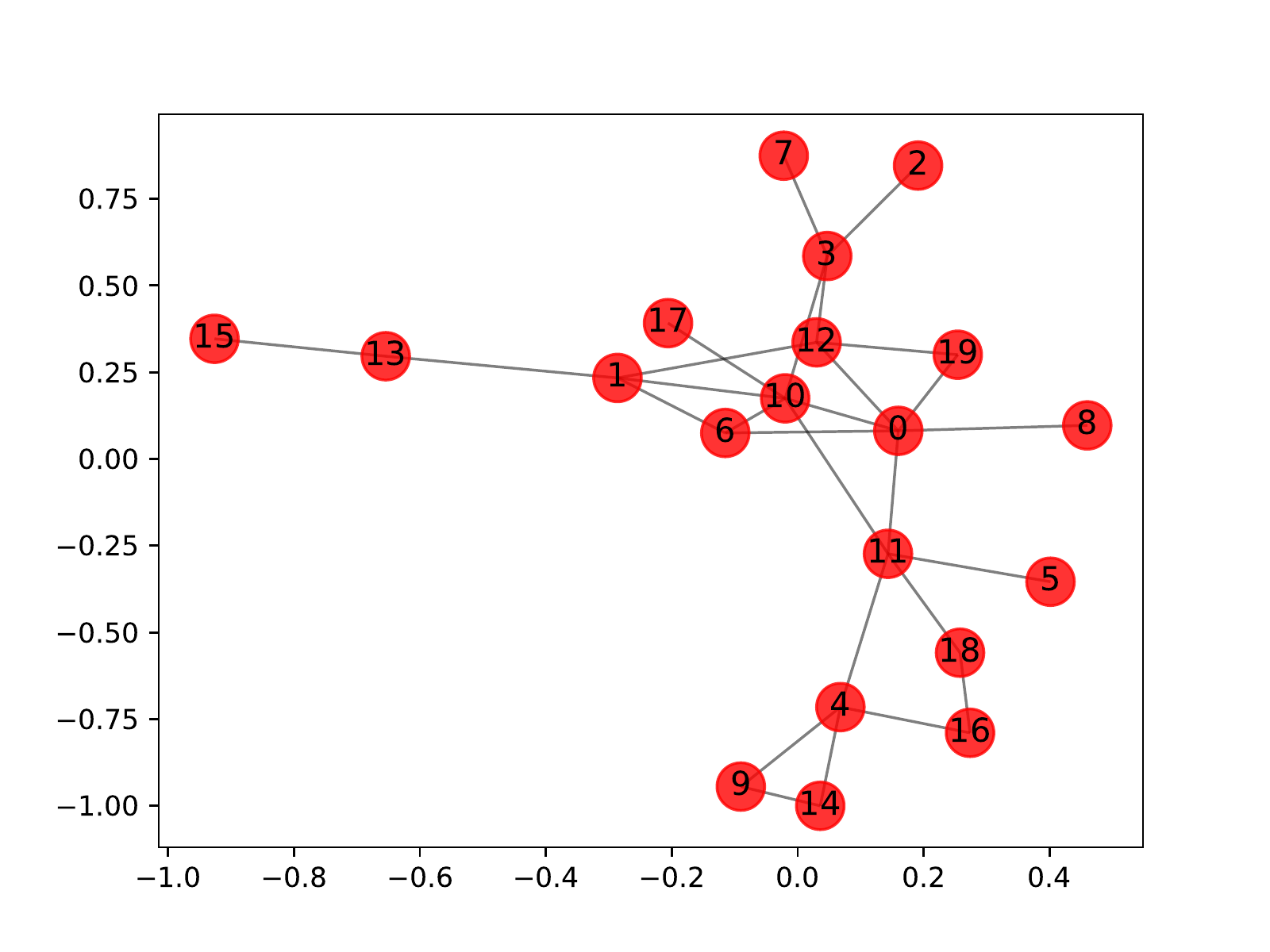}}\hspace{50pt}
	\caption{{\em Arbitary graph to a weakly chordal one}}
	\label{Sample3}
\end{figure}

\end{document}